\documentclass[aps,pre,floatfix]{revtex4}
\usepackage{graphicx}
\begin{document}

\title{Ideal Linear Chain Polymers with Fixed Angular Momentum}

\author{Matthew Brunner}
\author{J.M. Deutsch}
\affiliation{Department of Physics, UCSC}


\date{\today}

\begin{abstract}
The statistical mechanics of a linear non-interacting polymer chain with a
large number of monomers is considered with fixed angular momentum. The
radius of gyration for a linear polymer is derived exactly by functional
integration. This result is then compared to simulations done with a
large number of non-interacting rigid links at fixed angular momentum. The
simulation agrees with the theory up to finite size corrections.
The simulations are also used to investigate the anisotropic nature
of a spinning polymer. We find universal scaling of the polymer size
along the direction of the angular momentum, as a function of rescaled
angular momentum.
\end{abstract}


\maketitle

\section{Introduction}

Polymer chains in a vacuum have been shown to have great importance
in many experimental situations, most often in application to mass
spectroscopy techniques that utilize the desorption of proteins into a vacuum. The
ability to desorb very large molecules without compromising their
integrity has many uses in biology~\cite{Hillenkamp}. Understanding
the statistical mechanics of such systems may help to improve current
spectroscopic techniques by providing information about other quantities,
aside from mass, that may be usefully measured. Aside from other
applications~\cite{DeutschPolyVac}, there is also the intrinsic interest
in understanding such systems. 

The dynamics of vacuum polymers has been the subject of a number of recent
investigations~\cite{mossa,DeutschPolyVac,DeutschCerf,Taylor,DeutschExactVac}  and show
many unusual features different from what are seen in solution. Polymers close to the coil-globule
transition show small Lyapunov exponents~\cite{mossa}. However in many situations,
the dynamics are in accord with detailed microcanonical calculations,
implying that these systems are ergodic~\cite{Taylor}. Ideal chains show
highly oscillatory time correlation functions~\cite{DeutschPolyVac}, whereas
the addition of self interactions damps these oscillations although they
are still quite pronounced for small chains~\cite{Taylor}.  Although there
is no coupling to a heat-bath, the nonlinear dynamics of these models
gives rise to an effective damping of individual monomers which is similar
to that of Kelvin damping~\cite{SethnaBookKelvinFriction}.  Adding local angular 
potentials to polymers increases this type of damping but does not
inhibit oscillations for sufficiently long chains~\cite{DeutschCerf}.

In this paper, we further investigate the equilibrium statistics
of a noninteracting polymer chain of $N$ links in a vacuum, where
energy, momentum, and angular momentum are conserved. Previous
work~\cite{DeutschExactVac} found an analytical solution for a ring
polymer with conserved angular momentum.  The average radius of gyration
as a function of angular momentum was obtained. The most
surprising feature of that calculation is that the radius of gyration when
the angular momentum is zero is less than that of a chain without that
conservation law. Although the radius of gyration still is proportional
to $\sqrt{N}$ but now with a different proportionality constant.

The calculation performed for a ring chain~\cite{DeutschExactVac} does not
extend easily to linear chains.  Here we use an approach that diverges considerably
from the previous one to obtain the statistics of a linear chain under
the same conditions. In both cases, although the intermediate steps
produce rather lengthy expressions, the final answers are quite simple.
This suggests that there may be another underlying principle that can be
used to understand systems with constant angular momentum.

Aside from the average radius of gyration of such a system, there is
another important quantity that describes it, that has hitherto not
been studied. For finite angular momentum, the chain is expected to
become anisotropic, flattening in the direction of the angular momentum
vector. It is therefore of interest to calculate the average size
of a chain both perpendicular and parallel to the angular momentum.
However our analytic method is only applicable in calculating quantities
which have rotational symmetry such as the radius of gyration. It will
not work for studying chain anisotropy. Therefore we turn to numerical
methods to study this problem. We find, rather surprisingly, that the
chain continuously flattens in the direction of the angular momentum
showing universal scaling in the same rescaled variables used to
characterize the full radius of gyration.

\section{Exact Solution}
Following previous work ~\cite{laliena} we begin by writing down the entropy as a function of the total energy $E$, angular momentum $L$, and number of particles $N$,  including terms for the conservation laws to be enforced. 
\begin{equation}
W(E,L,N) = C \int \delta(E - K - \Phi)\delta^{(3)}(\mathbf{L} - \sum_i \mathbf{r}_i \times \mathbf{p}_i)\delta^{(3)}(\mathbf{r}_{cm})(\prod_{i=1}^N d^3r_id^3p_i)
\end{equation}
where $K$ is the total kinetic energy and $\Phi$ the total potential energy. We are choosing a coordinate system so that
the center of mass $\mathbf{r}_{cm} = 0$.

We take the potential energy to be the elastic energy of the polymer chain and in addition include an isotropic quadratic potential that
will be used to calculate the radius of gyration. In terms of the continuous arclength $s$ dependent position variables that we will use 
below, 
\begin{equation}\label{eq:DefPhi}
\beta \Phi =  \int^{Nl}_0 \frac{3}{2l} |\frac{d \mathbf{r}}{ds}|^2 + \epsilon r^2(s)  ds .
\end{equation}
where $l$ is the step length.

Because we are taking $N$ to be large, we can convert this to a partition function, with the energy conserving delta function expanded in exponential form.
\begin{equation}
Z(\beta,L,N) \propto  \int d^3k \int e^{ i\mathbf{k\cdot L} }e^{ -\beta(K+\Phi) }e^{ -i\mathbf{k}\cdot \sum_i \mathbf{r}_i\times \mathbf{p}_i }\delta^{(3)}(\mathbf{r}_{cm})(\prod_{i=1}^N d^3r_id^3p_i)
\end{equation}

After integrating over the momenta and taking into account rotational symmetry we arrive at the form~\cite{DeutschExactVac}

\begin{equation}\label{partzeta}
Z[\beta,L,N] = \frac{c}{L} \int_0^\infty \sin(kL)\zeta(\beta,k)dk
\end{equation}
where $\zeta$ is a function of the magnitude of $k$ and inverse temperature  $\beta$ only. The constant $c$ is of no physical importance and
\begin{equation}\label{rawzeta}
\zeta(\beta,k)= \int e^{ -\int^{Nl}_0 (\frac{Tmk^2}{2l} +\epsilon)(x(s)^2 + y(s)^2)+\epsilon z(s)^2 + \frac{3}{2l}  |\dot{\mathbf{r}}|^2  ds}\delta(\mathbf{r}_{cm}) \delta \mathbf{r}(s) .
\end{equation}
Now we perform a scaling on $L$, $k$, and $s$ in order to normalize out the length of the chain and the temperature.
\begin{eqnarray}
L^\prime = \frac{\sqrt{12}}{Nl\sqrt{mT}} L\\
k^\prime = \frac{Nl\sqrt{mT}}{\sqrt{12}} k\\
s^\prime = \frac{\sqrt{12}}{Nl} s
\end{eqnarray}
After doing these rescalings, the form of Eq. \ref{partzeta} remains unchanged, and Eq. \ref{rawzeta} changes to:
\begin{equation}\label{scaledzeta}
\zeta(\beta,k^\prime)= \int e^{-\int_0^{\sqrt{12}} (k^{\prime2}/2l + \epsilon^\prime)(x^\prime(s^\prime)^2 + y^\prime(s^\prime)^2 ) + \epsilon^\prime z^\prime(s^\prime)^2  + \frac{3}{2l} |\dot{\mathbf{r}^\prime}|^2 ds^\prime}\delta(\mathbf{r}^\prime_{cm}) \delta \mathbf{r^\prime}(s^\prime)
\end{equation}
It is interesting to note that the explicit dependence on $\beta$ has disappeared and $\zeta$ is now only a function of $k^\prime$. The dependence on the total length of the chain has also been factored into $k^\prime$, but the length of the individual monomers remains.
Now we expand the center of mass conserving delta function in complex exponential form.
\begin{equation}
\zeta(k^\prime)= \int e^{-\int_0^{\sqrt{12}} (k^{\prime2}/2l + \epsilon^\prime)(x^{\prime2} + y^{\prime2} ) + \epsilon^\prime z^{\prime2}  + \frac{3}{2l} |\dot{\mathbf{r}^\prime}|^2 + i x^\prime l_x + i y^\prime l_y + i z^\prime l_z ds^\prime}  dl_xdl_ydl_z \delta \mathbf{r^\prime}(s^\prime) .
\end{equation}
These terms can be combined with the existing exponents using completion of squares which results in path independent phase factors and a constant offset to the paths, e.g. $x^\prime (s) \to x^\prime (s) + const.$ . Since all possible paths are included in the integral, the offset has no effect.

\begin{figure}
\begin{center}
\includegraphics[width=\hsize]{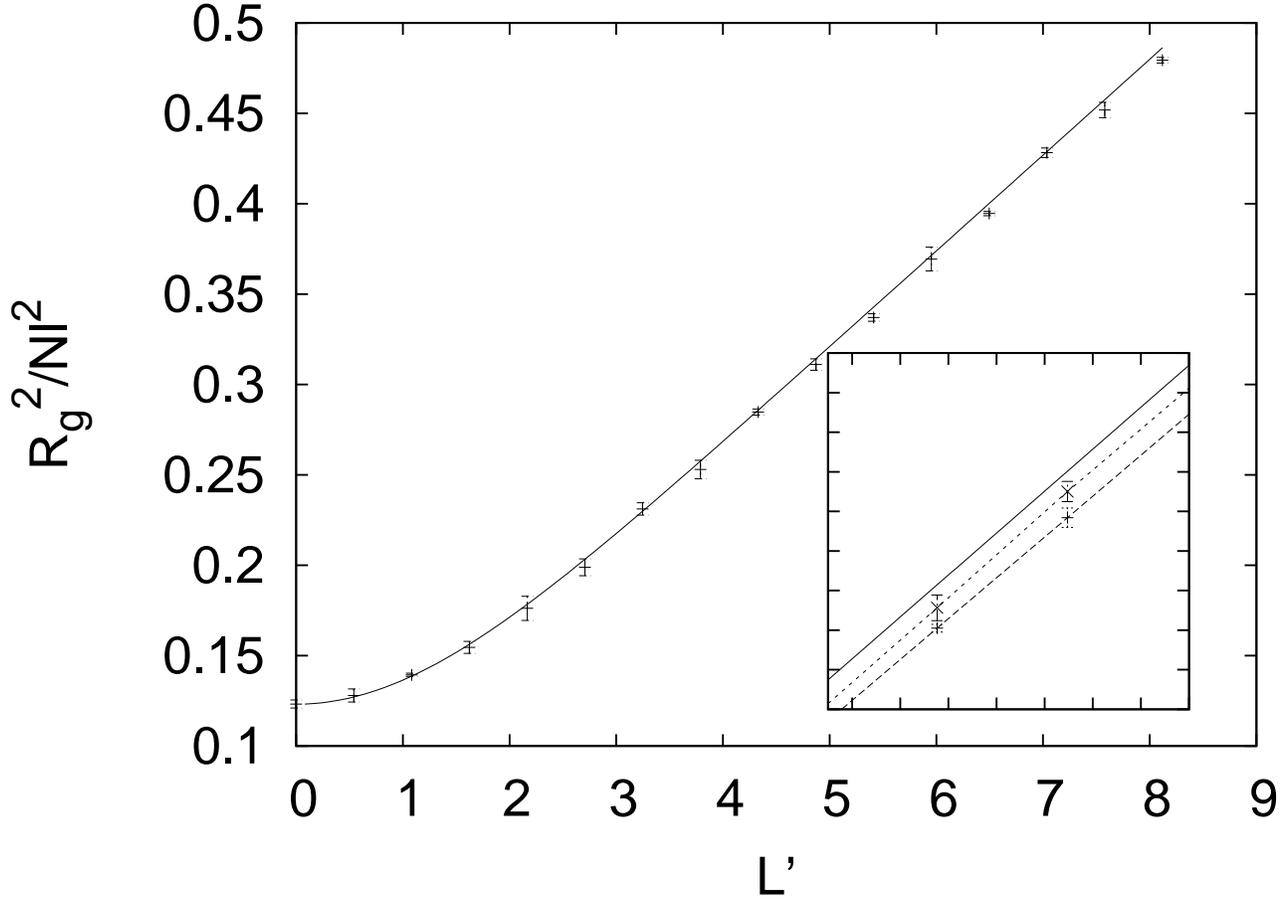}
\caption{Plot of $\frac{R_g^2}{Nl^2}$ along with simulation results
using a chain length $N = 128$. Inset is a section from $L^\prime=5.5$
to $L^\prime=7$ showing the exact solution as a solid line along with results from simulations done with $N = 32$ (farthest from line)
and $N = 64$. }
\label{fig:rgplot}
\end{center}
\end{figure}

We now have a nice separation into a simple Gaussian integral and a factor that is very similar to the path integral form of a harmonic oscillator. 
\begin{equation}
(\int e^{\sqrt{12}[ -(l_x^2 + l_y^2)/4(\frac{k^{\prime2}}{2l} +\epsilon^\prime) - l_z^2/4\epsilon^\prime]}dl_xdl_ydl_z)  (\int e^{-\int_0^{\sqrt{12}} (k^{\prime2}/2l + \epsilon^\prime)(x^{\prime2} + y^{\prime2} ) + \epsilon^\prime z^{\prime2}  + \frac{3}{2l} |\dot{\mathbf{r}^\prime}|^2 } \delta \mathbf{r}^\prime)
\end{equation}
The first integrals of the $l$'s are trivially integrated but we must still perform the path integral in parentheses. 
These decouple into path integrals over the $x$, $y$, and $z$ directions and each one has an action of the form
\begin{equation}
\int_0^{\beta_0} \frac{M}{2}(\dot{x}^2 + \omega_0^2 x^2)dt .
\end{equation}
$M$ does not depend on $k$ or $\epsilon$, and $\omega_z$ has no $k$ dependence.
After making this analogy, we can see that for any two fixed path endpoints, the functional integration will give the Greens function for a thermal harmonic oscillator. To obtain a sum over all paths, we simply need to integrate the Greens function $\rho$ ~\cite{feynman} over both its arguments.
\begin{equation}
\zeta_i = \int \rho_i(x_1,x_2)dx_1dx_2
\end{equation}
where $i = 1,2,3$ represent the $x$, $y$, and $z$, path integrals respectively.
The Greens function is easily integrable and the result is 
\begin{equation}\label{eqn:zomega}
\zeta_i \propto \frac{1}{\sqrt{\omega_i \cosh{(\sqrt{12}\omega_i)}}}
\end{equation}
Putting all of this back into Eq. \ref{partzeta} we get the form for $Z$
\begin{equation}
Z(L^\prime) = \frac{c}{L} \sqrt{\frac{\omega_z}{\sinh{(\sqrt{12}\omega_z)}}} \int_0^\infty k^\prime \sin(k^\prime L^\prime) \frac{\omega}{\sinh(\sqrt{12}\omega)} dk^\prime .
\end{equation}
The radius of gyration can be expressed as
\begin{equation}
R_g^2 = -\frac{1}{Nl}\frac{\partial \ln Z}{\partial \epsilon}\vert_{\epsilon=0}
\end{equation}
The logarithmic derivative allows us to separate the factors in $Z$ and treat each as terms in a sum.
The first term is simple
\begin{equation}
\lim_{\epsilon\to0}\frac{\partial \ln}{\partial \epsilon}\sqrt{\frac{\omega_z}{\sinh{(\sqrt{12}\omega_z)}}} = -\frac{1}{18}N^2l^3
\end{equation}
The second term is more complex due to the integral, so we begin by rewriting the logarithmic derivative as a ratio.
\begin{equation}
\lim_{\epsilon\to0} \frac{\frac{\partial}{\partial \epsilon}\int_0^\infty k^\prime \sin(k^\prime L^\prime) \frac{\omega}{\sinh(\sqrt{12}\omega)} dk^\prime}{\int_0^\infty k^\prime \sin(k^\prime L^\prime) \frac{\omega}{\sinh(\sqrt{12}\omega)} dk^\prime}
\end{equation}
Since the integral is over $k^\prime$ we can pull the derivative and limit inside the integration. This simplifies the integrand considerably as $\lim_{\epsilon\to0} \omega = k^\prime/\sqrt{3}$
\begin{equation}
\lim_{\epsilon\to0} \frac{\frac{\partial}{\partial \epsilon}\int_0^\infty k^\prime \sin(k^\prime L^\prime) \frac{\omega}{\sinh(\sqrt{12}\omega)} dk^\prime}{\int_0^\infty k^\prime \sin(k^\prime L^\prime) \frac{\omega}{\sinh(\sqrt{12}\omega)} dk^\prime}  = \frac{L^\prime N^2}{6\pi l^3\tanh(\frac{L^\prime \pi}{4}) }
\end{equation}
Put all of the pieces together to obtain our final result.
\begin{equation}
\label{eq:finalresult}
\frac{R_g^2}{Nl^2} = \frac{1}{18} + \frac{L^\prime}{6\pi\tanh(\frac{L^\prime\pi}{4})}
\end{equation}
A plot of this is shown in Fig. \ref{fig:rgplot} by the solid line. The simulation results also displayed will be discussed in Sec. \ref{sec:simresults}.
Note that for $L^\prime = 0$, $R_g^2/(N l^2) = (1/18) + 2/(3 \pi^2) \approx 0.1231$, which is about $74\%$ of $1/6$, the value of the same
quantity without conservation of anguar momentum enforced.

For the same reason as for the ring calculation~\cite{DeutschExactVac}, for fixed $L^\prime$ and $N\to \infty$, we are permitted to take this
system to be at constant temperature, because as was discussed in detail~\cite{DeutschExactVac}, the relation between energy and temperature
has no dependence on $L^\prime$ in this limit.

\section{High $L$ limit}
In the limit of large $L$, Eq. \ref{eq:finalresult} gives $R_g^2/Nl^2 \to L^\prime/6\pi$. One would expect typical configurations at high $L$ would consist of a fairly straight line rotating rapidly around its center. The solution for the equilibrium case of constant angular velocity can be found by setting the change in tension along the chain equal to the centripetal force. This gives an equation for the position of each monomer as a function of its location along the chain, $r(s)$, for simplicity defined with $s=0$ at the center of the chain. For large $N$, the tension approaches zero at the ends of the chain.
\begin{equation}
-k\frac{d^2r}{ds^2} = m \omega^2 r \\
\end{equation}
where $k$ is the entropic elastic spring coefficient.
From this we get that $r(s) \propto \sin(\sqrt{\frac{m}{k}}\omega s)$, and that $Nl\omega/2\sqrt{k} = \pi/2$. Higher modes exist, but they double back and are expected to not be minimal free energy solutions. They are also not the straight chain model we are assuming. Using the integral definition of angular momentum
\begin{equation}
L = \frac{m \omega}{l}\int_0^{Nl}  r^2(s)ds 
\end{equation}
the radius of gyration squared is
\begin{equation}
R_g^2 \equiv \frac{1}{Nl}\int r^2(s)ds = \frac{1}{\omega m N } L = \frac{1}{\sqrt{mk}\pi}L = \frac{l}{\sqrt{3mT}\pi}L = \frac{L^\prime Nl^2}{6\pi}
\end{equation}
And we see that in the limit of high $L$, our model behaves as one would expect.
\section{Partition Function}
\begin{figure}
\begin{center}
\includegraphics[width=\hsize]{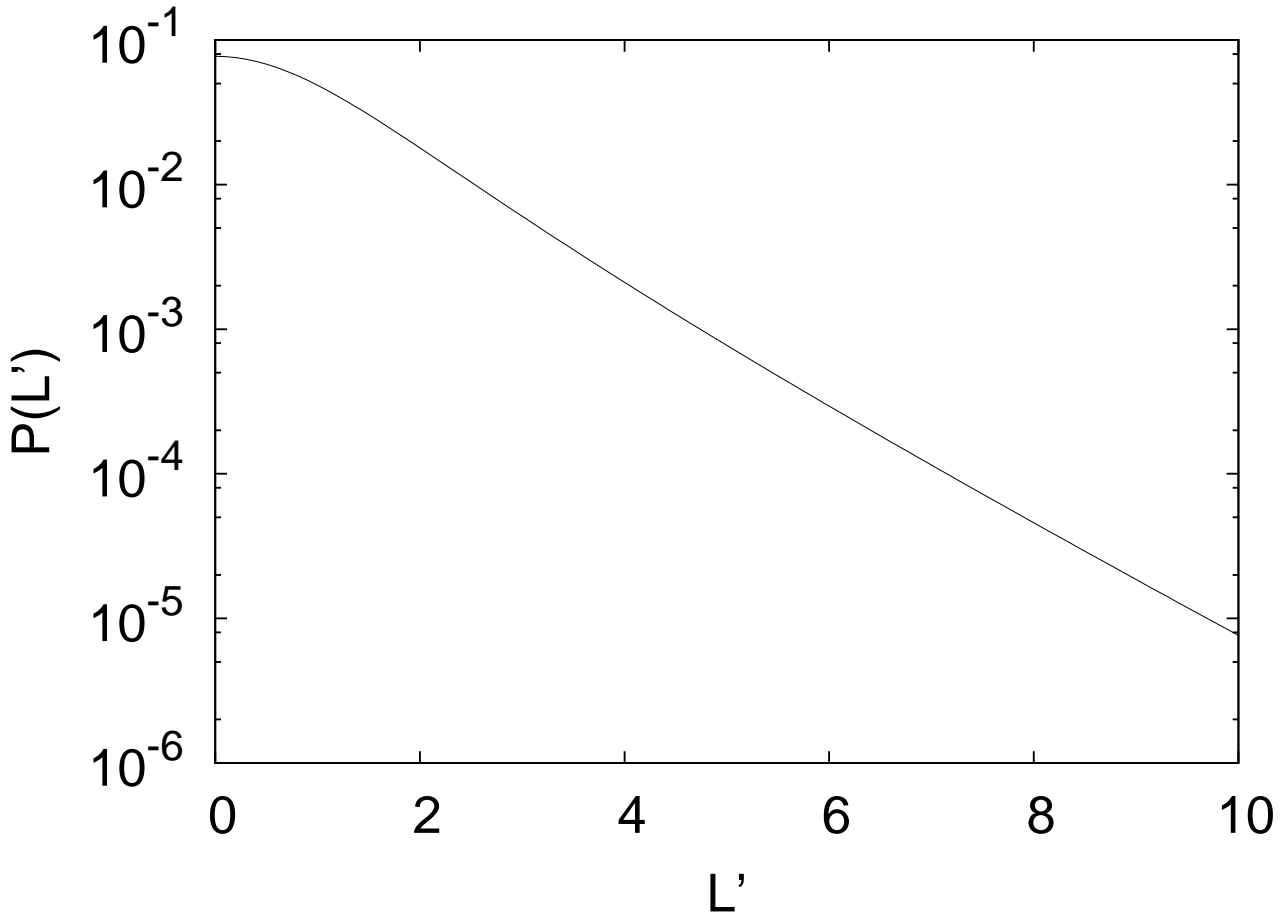}
\caption{Plot of $P(L^\prime))$, with clear non-Gaussian behavior for high $L^\prime$}
\label{fig:P}
\end{center}
\end{figure}

The partition function itself can be obtained easily from Eq. \ref{eqn:zomega} by taking $\epsilon \to 0$ and performing the integration.
\begin{equation}
\lim_{\epsilon \to 0} Z = \frac{1}{L} \int k^{\prime2} \frac{\sin(k^\prime L^\prime)}{\sinh(2k^\prime)}dk^\prime = \frac{\pi^3}{32L^\prime}\frac{\tanh(L^\prime \frac{\pi}{4})}{\cosh^2(L^\prime \frac{\pi}{4})}
\end{equation}
To obtain the distribution over $L^\prime$ we need to normalize this Z, taking in to account the angular integration over the direction of $L^\prime$.
\begin{equation}
\int 4\pi L^{\prime2} Z dL^\prime = \pi^2
\end{equation}
so our normalized partition function is 
\begin{equation}
Z = \frac{\pi}{32L^\prime}\frac{\tanh(L^\prime \frac{\pi}{4})}{\cosh^2(L^\prime \frac{\pi}{4})}
\end{equation}
A plot of this is displayed in Fig. \ref{fig:P}
The distribution shows a long exponential tail into high $L^\prime$.

\section{Simulation Results}
\label{sec:simresults}
In addition to the exact calculation, further studies were carried out using simulation. First the analytical results concerning the radius of gyration as a function of $L\prime$ were verified. Second we were able to investigate other quantities that were beyond the means of our analytic method.

\begin{figure}
\begin{center}
\includegraphics[width=\hsize]{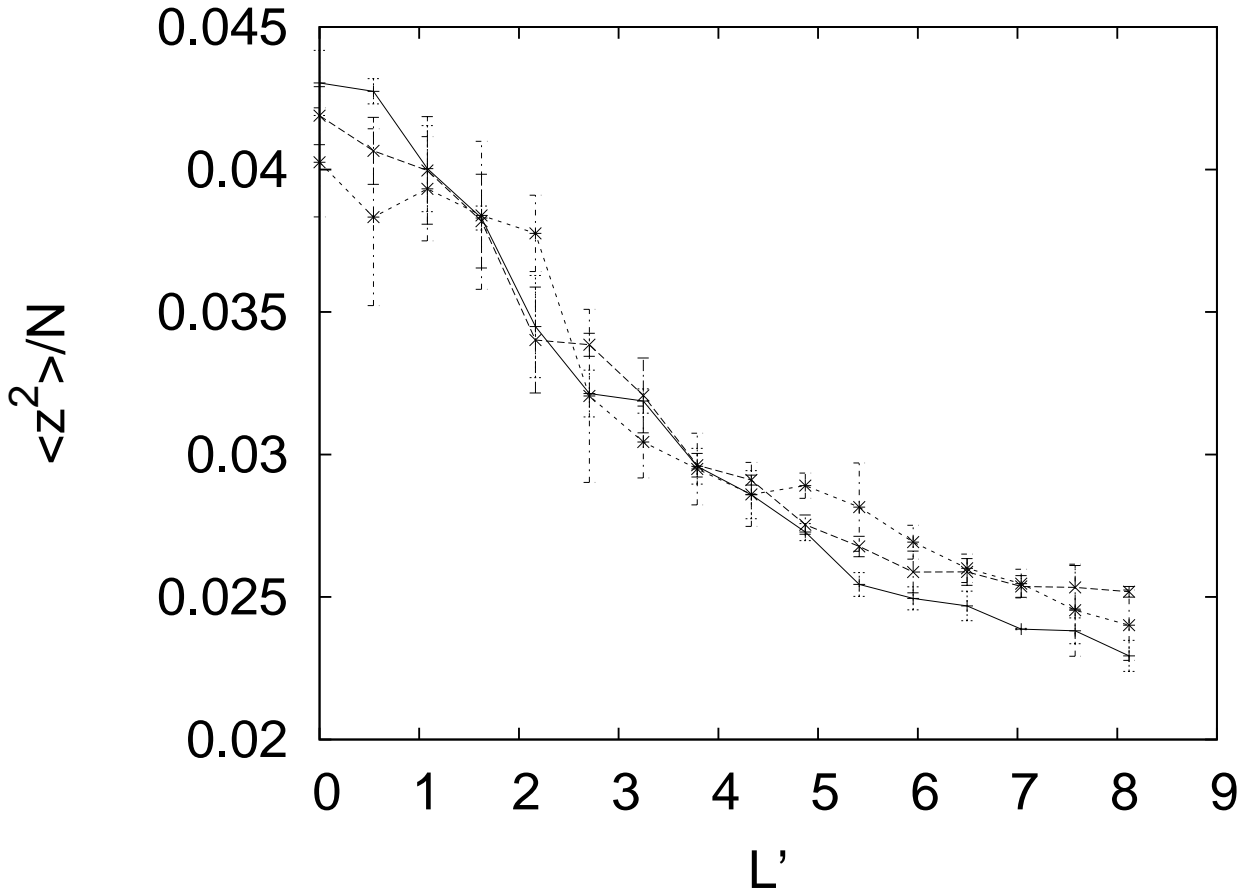}
\caption{Plot of radius of gyration in the direction of angular momentum as a function of rescaled $L$.}
\label{fig:zsq}
\end{center}
\end{figure}

We used a molecular dynamics method ~\cite{DeutschCerf} that was developed to simulate chains in a vacuum. It
consisted of freely rotating rigid links, conserving energy and angular
momentum. Rigid links were employed to minimize problems with equilibration that are often seen with one dimensional nonlinear systems~\cite{FPU,BermanIzrailev}. The input angular momentum and the output measurements were
scaled by the total chain length for comparison. The initial values of
the angular momentum were chosen to cover a range of $L^\prime$ values, and the
angular momentum was explicitly checked and conserved during the runs.
First the simulation was compared to the theoretical results for the
normalized radius of gyration. The results of these simulations can be
seen in Fig. \ref{fig:rgplot}, and show excellent agreement with the
theoretical prediction. The main plot shows the exact result (solid line)
along with data for chains with $N=128$. The inset shows that the deviation in the asymptotic form for high
$L^\prime$ decreases as the number of simulation chain links is increased and
appears to be due to the finite size of the simulation system. The two chains lengths
used in the inset are $N=32$ and $N=64$ which are more accurate than the data
for $N=128$. They also indicate that the finite size corrections to the
analytic form are $O(1/N)$.

For nonzero angular momentum the polymer chain is expected to become
anisotropic. The analytic methods used earlier require an isotropic
form for the quantities being averaged, causing the investigation
of the anisotropy by such means to be outside the scope of our
analysis. Therefore we turned to the simulation to explore the aspect
ratio of the polymer as a function of $L^\prime$  The aspect ratio is defined
as $\sqrt{\langle z^2 \rangle /\langle R_g^2 \rangle}$.The aspect ratio
itself is dominated by the asymptotic linear behavior of the radius of
gyration, and displays what appears to be an inverse power law falloff
as expected. More interestingly, $\langle z^2 \rangle$ itself appears to
fall off as $L^\prime$ increases.  This is shown in Fig. \ref{fig:zsq}  where the
vertical axis is scaled in the same manner as for the radius of gyration,
by dividing by $\frac{N}{\sqrt{12}}$ and taking $l,m,T$ all to be unity, and the horizontal axis is $L^\prime$  The three chain
lengths seem to collapse onto one curve after the rescaling of length
and angular momentum. This is unexpected because for a ideal gaussian
chain each dimension has independent statistics. In the limit of high L
the rigid link model should approach a straight line solution. However
this is unlikely to be an explanation, as the straight line regime
would imply a leveling off of the radius of gyration, which is not
seen. Moreover, the falloff is independent of the number of chain links
in the simulation and collapsed onto a single curve, so is not likely
due to the non gaussian nature of the model. This decrease is slow and
does not appear to have a nonzero asymptote, and is likely to be a low
power law or logarithmic in nature.


\section{Conclusions}

In this paper, we considered the equilibrium properties of an ideal linear chain in a vacuum
that conserves energy, momentum, and angular momentum. We were able to compute the average
radius of gyration of such a chain as a function of its angular momentum $L$. We also computed
the distribution of angular momenta for chains in thermal equilibrium. 
We verified that our analytical result for the radius of gyration is correct by performing
numerical simulation and by analyzing its asymptotic form in the limit of large angular momentum.
The derivation of this result differs from that of a ring chain but in both cases the final
result is relatively simple involving hyperbolic trigonometric functions. The underlying reason
for this is still unclear.

Our numerical simulations show that the radius of gyration perpendicular
to the angular momentum vector increases with $L$, as to be expected,
however in addition to this, the radius of gyration parallel to the
angular momentum, which we take to be in the $z$ direction, decreases.
This is very different than what a naive analysis would suggest. If we
were to go to a frame rotating at angular velocity $\Omega$, then in
thermal equilibrium (without angular momentum conservation), this is
equivalent to an additional potential $- m \Omega^2 r^2_{xy}/2$, where
$r_{xy}$ is the projection of a coordinate onto the $x-y$ plane.  For an ideal
gaussian chain, all three directions decouple and$\langle z^2\rangle$
is independent of $L$. However a more rigorous analysis based on the
approach of this paper is not equivalent to this, and it is not clear
how statistics in the $x-y$ plane and $z$ direction are coupled to cause
this effect.




\begin{thebibliography}{}
\bibitem{Hillenkamp} F. Hillenkamp (Editor), J. Peter-Katalinic (eds.). 
\bibitem{DeutschPolyVac} J.M. Deutsch, Phys. Rev. Lett. {\bf 99}, 238301 (2007).
\bibitem{mossa} A. Mossa, M. Pettini, and C. Clementi, Phys. Rev. E {\bf 74} 041805 (2006).

\bibitem{DeutschCerf} J.M. Deutsch, arXiv:1003.0944v2 
\bibitem{Taylor} M. P. Taylor, K. Isik and J. Luettmer-Strathmann,  Phys. Rev. E {\bf 78}, 051805  (2008).
\bibitem{DeutschExactVac} J.M. Deutsch, Phys. Rev. E {\bf 77}, 051804 (2008).
\bibitem{laliena} V Laliena, Phys. Rev. E {\bf 59}, 4786 (1999).
\bibitem{feynman} R. P. Feynman, Statistical Mechanics: A Set of Lectures, Westview Press; 2nd Edition (1998)
\bibitem{SethnaBookKelvinFriction} J.P. Sethna ``Statistical Mechanics Entropy, Order
\bibitem{DeGennesBook} P.G. de Gennes ``Scaling Concepts in Polymer Physics" Cornell University Press (1985).
\bibitem{FPU} E. Fermi, J. Pasta, and S. Ulam, {\em Studies of nonlinear problems} 
(Los Alamos Document LA-1940, 1955).
\bibitem{BermanIzrailev} For a review see G.P. Berman and F.M. Izrailev, Chaos
\end{thebibliography}
\end{document}